\documentclass[final,aps,prl,twocolumn,superscriptaddress,groupedaddress,nofootinbib,floatfix,x11names]{revtex4-1}
\usepackage{units}
\usepackage[normalem]{ulem}
\usepackage{times}
\usepackage{amsmath}
\usepackage{amssymb}
\usepackage{bbold}
\usepackage{paralist}
\usepackage{xcolor}
\usepackage{color}
\usepackage{slashed}
\usepackage{graphicx} 
\usepackage{xspace}
\usepackage{siunitx}
\newcommand{\myparagraph}[1]{\bigskip \noindent\textbf{#1}}
\newcommand{\CO}{\mathcal{O}}
\newcommand{\dec}{\text{dec}}
\newcommand{\esc}{\text{esc}}
\newcommand{\flip}{\text{flip}}
\newcommand{\bmax}{b_{\text{max}}}
\newcommand{\GN}{G_{\text{N}}}
\usepackage[margin=false,inline=true,index=true,draft]{fixme}
\fxsetup{inlineface=\footnotesize}
\fxusetheme{colorsig}
\FXRegisterAuthor{oleg}{anoleg}{O.R.}

\def \be {\begin{equation}}
\def \ee {\end{equation}}
\def \bea {\begin{eqnarray*}}
\def \eea {\end{eqnarray*}}

\def \cnb {C$\nu$B\xspace}
\definecolor{darkred}{RGB}{166,42,23}

\usepackage[colorlinks,bookmarks=false,linkcolor=blue,urlcolor=blue,citecolor=blue]{hyperref}

\begin{document}

\title{Lepton number survival in the Cosmic Neutrino Background}

\author{Oleg~Ruchayskiy}
\affiliation{Niels Bohr Institute, Copenhagen University, Blegdamsvej 17, DK-2100 Copenhagen, Denmark}

\author{Vsevolod Syvolap}
\affiliation{Lorenz Institute, Leiden University, Niels Bohrweg 2, NL-2333 CA,  Leiden, The Netherlands}

\author{Robin~W\"ursch}
\affiliation{\'Ecole Polytechnique F\'ed\'erale de Lausanne, FSB/ITP/LPPC, BSP, CH-1015, Lausanne, Switzerland}
\date{\today}

\begin{abstract}
  The Hot Big Bang model predicts the existence of a \emph{cosmic neutrino background}.
  The number of particles and anti-particles in this primordial bath of neutrinos can be different -- a memory of processes that took place at very early epochs.
  If neutrinos were massless, this asymmetry would not change once neutrinos froze out.
  However, in the case of massive particles, the asymmetry is not protected by conservation laws and can get erased via helicity-flipping scatterings off matter inhomogeneities.
  We evaluate this helicity-flipping rate and demonstrate that if relic lepton asymmetry ever existed,  it would remained largely intact in the Earth's neighborhood for realistic values of neutrino masses.
\end{abstract}


\maketitle

\myparagraph{Introduction.} The Hot Big Bang theory predicts that along with the Cosmic Microwave background there exists a bath of primordial neutrinos -- the \emph{Cosmic Neutrino background} (\cnb). The neutrinos decoupled from the thermalized primordial plasma at temperatures $T_{\dec} \sim$ few MeV (see e.g.~\cite{dolgov_cosmology_2008}) and their temperature today is predicted to be 
$T_\nu \simeq (4/11)^{1/3}T_{\rm cmb}$  (see e.g.~\cite{Dolgov:97sm}).
At decoupling, neutrinos had a relativistic Fermi-Dirac distribution, $f_{\rm FD}(p)$, with the temperature $T_{\dec}$ (up to small corrections~\cite{Dolgov:98addendum}).
The distribution function of decoupled neutrinos subsequently evolves (neglecting inhomogeneities) as
\begin{equation}
  f_\nu(p,t) = f_{\text{\sc fd}}\left(p \frac{a(t)}{a_\dec}\right)
  \label{eq:1}
\end{equation}
conserving its shape in terms of conformal momentum.
Possibilities of direct detection of \cnb have been discussed since the 1960s~\cite{Weinberg:1962zza,Stodolsky:1974aq} (see~\cite{ringwald_prospects_2009} for review), but recent years saw a surge of interest to such type of experiments, thanks to technological advances \cite{Duda:2001hd, Gelmini:2004hg, blennow_prospects_2008,cocco_probing_2008, Ringwald:2009bg, Faessler:2011qj,Betts:2013uya,Faessler:2013jla,Birrell:2014qna,Long:2014zva, Yoshimura:2014hfa,Zhang:2015wua,Vogel:2015vfa,Faessler:2016tjf,Li:2016qsu, Domcke:2017aqj,PTOLEMY:2018jst,PTOLEMY:2019hkd,Akhmedov:2019oxm,Mikulenko:2021ydo, Bauer:2021uyj}.
The measurement of the \cnb would confirm one of the central predictions of the Hot Big Bang model and pave the road to future measurements of \emph{anisotropies of \cnb}~\cite{Michney:2006mk,Hannestad:2009xu,Lisanti:2014pqa}.
It would also open a window to new physics in the neutrino sector~\cite{Davidson:2003ha,Mangano:2006ar,Basboll:2009qz,Fuller:2011qy,Diaz:2015aua}.
In particular, the \cnb can be hiding a large \emph{relic lepton number}.\footnote{By ``\emph{lepton number}'' in this paper we always mean \emph{total lepton number}.}

Indeed, the existing upper bounds~\cite[see e.g.][]{Lesgourgues:99,Serpico:05,Pastor:2008ti,Mangano:10, KATRIN:2022kkv} or recent hints of detection \cite{Matsumoto:2022tlr,Escudero:2022okz}
admit a lepton asymmetry as large as $|\eta_L|\lesssim \mathcal{O}(10^{-1})$.\footnote{Lepton asymmetry $\eta_L$ is defined as $(n_L - n_{\bar L})/{s}$, where $n_L$ ($n_{\bar L}$) is the total number density of leptons $L$ (anti-leptons $\bar L$), and $s$ is the entropy.}

This asymmetry can in theory be measured via   \textit{e.g.}\ the \emph{Stodolski effect} \cite{Stodolsky:1974aq,Hagmann:1999kf,Long:2014zva,Akhmedov:2019oxm}, although the detection threshold is beyond the reach of current technologies \cite{Duda:2001hd}.

The measurement of the relic neutrino asymmetry could provide information about leptogenesis models~\cite{Shaposhnikov:08a,Gu:2010dg,Eijima:2017anv,Eijima:2020shs} or about other beyond-the-Standard Model processes taking place in the early Universe~\cite{Foot:1995qk,Enqvist:1996eu,Abazajian:2004aj,Solaguren-Beascoa:2012wpm,Zhang:2015wua,Chen:2015dka, Gelmini:2020ekg,Escudero:2022okz,Domcke:2022uue,Burns:2022hkq,Kawasaki:2022hvx}.

If neutrinos were massless, the lepton asymmetry, stored in the neutrino sector would remain unchanged after decoupling.
However, the existence of neutrino masses means that this asymmetry changes via \emph{helicity-flipping} gravitational scattering of neutrinos of inhomogeneities.
In the long run, such processes should fully erase any left-over lepton asymmetry.

\emph{The goal of this paper} is to determine the rate at which such an erasure happens.
We will find that for admissible values of neutrino masses, only a small fraction of the neutrino population may undergo helicity-flipping until the present age of the Universe and the total lepton number would remain hidden in the neutrino background.

\bigskip

The paper is organized as follows. First, we define lepton asymmetry for Dirac or Majorana neutrinos.
Next, we argue that only a small fraction of neutrinos in the Earth's vicinity are gravitationally bound to the Milky Way (and thus would have their lepton number erased).
After that, we compute the helicity-flipping cross-section for neutrinos scattering off matter inhomogeneities.
We then estimate the helicity-flipping rate and show that over the history of the Universe most neutrinos have never experienced such a process.
We conclude with discussion of potential observability of the relic lepton number.

\myparagraph{Lepton asymmetry for Dirac and Majorana neutrinos.}
In the Standard Model with massless neutrinos, there are three \emph{classically conserved} flavor lepton numbers (asymmetries between leptons and anti-leptons of a given generation). 
Neutrino flavor oscillations redistribute these asymmetries between flavors while leaving the total lepton number unchanged.
Suppose neutrino masses are of Majorana type (we treat Majorana masses as coming from the Weinberg operator \cite{Weinberg:1980bf} leaving aside its microscopic origin).
In that case, the conserved lepton number cannot be defined.
However, we can think of \emph{left-helical} (LH) and \emph{right-helical} (RH) neutrino states as neutrinos and anti-neutrinos correspondingly. 
The \emph{lepton asymmetry} is then simply a disbalance between LH and RH states. 

If neutrino mass is of the Dirac type, the total lepton number is of course conserved but is redistributed among four Dirac spin states -- two from the Standard model sector and two ``sterile'' (right-handed) counterparts.
The same helicity-flipping processes then equilibrate left-chiral (active) and right-chiral (sterile) states.
The LH-RH asymmetry then means that sterile particle and sterile anti-particle states are populated at different rates, leading to the change of the mean helicity \emph{in the active sector}.
As we will see below, the computations in both cases are similar up to trivial numerical factors.

\begin{figure}
  \includegraphics[width=\linewidth]{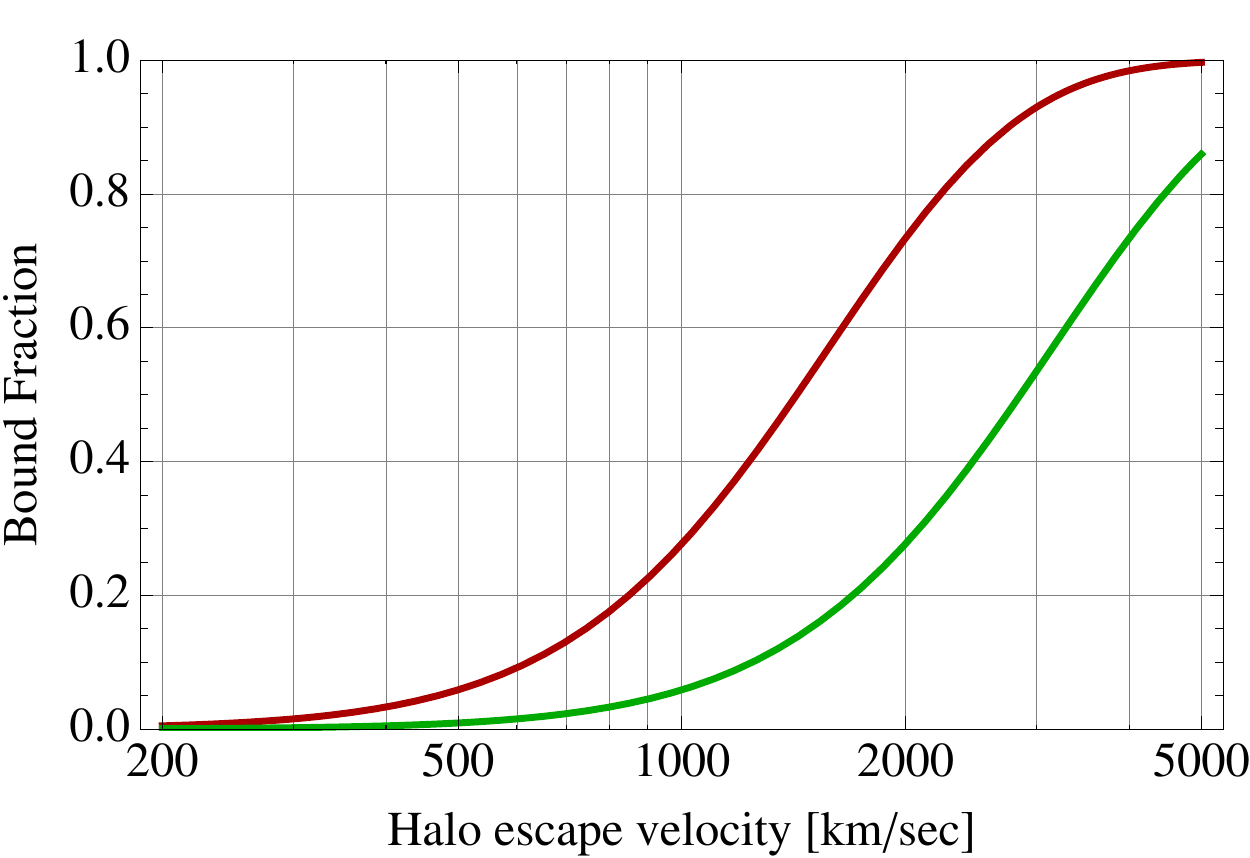}
  \caption{Fraction of neutrinos with velocity below $v_\esc$ for 2 different mass eigenstates: $m = 0.05 $~eV green line and $m = 0.1$~eV red line. The approximation $p = mv$ was used for Fermi distribution with temperature $T_\nu = 1.9$ K.}
  \label{fig:fraction}
\end{figure}

\myparagraph{Fraction of gravitationally bound neutrinos.} Neutrinos that are gravitationally bound to stars, galaxies, etc., change directions of their momenta but not their spins. Therefore any helicity imbalance equilibrates after a few orbital times.  
We estimate the bound fraction by computing the number of neutrinos whose velocity $v$ is below the escape velocity of an object~$v_{\esc}$:
\begin{equation}
 \label{eq:fraction}
 F(v_\esc) \equiv n_{\rm tot}^{-1}\int_0^{v_{\esc}}d^3v\, f_\nu(v),
\end{equation}
where $f_\nu(v)$ today is given by
\begin{equation}
    \label{eq:FD}
    f_\nu(v) = \frac1{\exp\left(\frac{mv -\mu}{T_\nu}\right) + 1}.
\end{equation}
Here $m$ is the heaviest neutrino mass,\footnote{It will dominate helicity-flipping rate.} $T_\nu\simeq \SI{1.9}{K}$ is the \cnb temperature today, $\mu$ is the chemical potential, $n_{\rm tot}$ is the normalisation, ensuring that $F(v) \to 1$ as $v \to 1$.\footnote{We work in natural units, $c=k_B = \hbar = 1$.  } In what follows, $\mu_\nu/T$ is assumed to be small.
The function $F(v_\esc)$ is presented in Fig.~\ref{fig:fraction} for $m = \SI{0.05}{eV}$ and $\SI{0.1}{eV}$.\footnote{We adopt here two reference values of the neutrino mass: $m=\SI{50}{meV}$ and $m = \SI{100}{meV}$. In the $\Lambda$CDM model extended by neutrino masses alone the sum of the neutrino masses is limited to $\sum m_\nu < \SI{129}{meV}$ when combining the Planck measurements \cite{Planck:2018vyg} with those of eBOSS \cite{eBOSS:2020yzd}. The bound shrinks down to $\SI{100}{meV}$ if more datasets are combined \cite{eBOSS:2020yzd}.}

The Milky Way (MW) has $v_{\esc} \approx 500-600$~km/s~\cite{Smith:2006ym,Kafle:2014xfa} and the fraction of bound neutrinos obeying  distribution~\eqref{eq:FD} is $\sim 9\%$ for $m=\SI{0.1}{eV}$ (and $\sim 1\%$ for $m=0.05$~eV neutrinos). 
These numbers should be further corrected for local overdensity of neutrinos, $\delta_\odot$, see e.g.~\cite{Ringwald:2004np,loverde_neutrino_2014,deSalas:2017wtt,Zhang:2017ljh,deSalas:2019kpa,Mertsch:2019qjv}. 
Its estimates depend on the assumed mass distribution in the Milky Way.
The most recent work \cite{Mertsch:2019qjv} reports overdensities $\delta_\odot \simeq 12\%$ for $m = \SI{0.06}{eV}$ and $\delta_\odot \simeq 50\%$ for $m = \SI{0.1}{eV}$, including effects of the Milky Way, Andromeda galaxy and the Virgo cluster.
The resulting fraction of bound neutrinos thus does not exceed $9\times 1.5 \approx 13.5\%$ for $m\simeq \SI{0.1}{eV}$ (and is of the order of $2\%$ for $m \simeq \SI{0.05}{eV}$).

\myparagraph{The gravitational helicity-flip rate.} The computation of the helicity-flipping rate is similar to the well-known Rutherford scattering computation. 
The main complication comes from the expanding Universe as the characteristic scattering time for the largest structures is larger than the Hubble time.
We will bypass this complication by estimating the helicity-flipping rate from above, using an auxiliary computation in the Minkowski space with a small overdensity. 

We start by considering  Dirac neutrinos with the mass $m$. A small perturbation over the Minkowski metric  due to a point-mass $M$ is described via metric $h_{\mu\nu}$
\begin{equation}
\label{eq:metric}
g_{\mu\nu} = \eta_{\mu\nu} + h_{\mu\nu} \qquad \mathrm{with}\,\, |h_{\mu\nu}| \ll 1\;,
\end{equation}
or, correspondingly, vierbein,
\begin{equation}
    g_{\mu \nu} = e^a_\mu e^b_\nu \eta_{a b},\quad e^a_\mu = \delta^a_\mu + \frac{h^a_\mu}{2}
\end{equation}
Writing the Dirac equation in the metric $g_{\mu\nu}$ and expanding to first order in $h$ we  arrive at the perturbed Dirac equation
\begin{equation}
\label{eq:dirac_final}
	i\gamma^a\partial_a\psi -m \psi = \frac{i}{2}h^{ab}\gamma_a\partial_b\psi,
\end{equation} 
where indexes $a, b$ correspond to flat space-time metrics. 

The $S$-matrix element is given by
\begin{equation}
\label{eq:smatrix}
	 S_{fi} = \frac{1}{2}\int d^4x\bar{\psi}_f(x)h^{ab}(x)\gamma_a\partial_b\psi_i(x)
\end{equation}
which leads to the differential cross-section for the helicity-flipping process (see~\cite{MSc_Robin_Wursch} for details of the computation):
\begin{equation}
  \label{eq:flip_dif_crosssection}
  \frac{d\sigma}{d\Omega} = \frac{(\GN Mm)^2}{16p^4\sin^4\frac\theta2}E^2\left(1-\cos\theta\right)
\end{equation}
where $\GN$ is the Newton's constant, $E$ and $p = |\vec p|$ are energy and momentum of the neutrino; $\theta$ is the scattering angle.\footnote{Eq.~\eqref{eq:flip_dif_crosssection} agrees with the computations of Ref.~\cite{Aldrovandi:1994es}. The angular dependence also agrees with \cite{choudhury_gravitational_1989} although the latter has the prefactor with the wrong dimensionality.} 

The total helicity-flipping cross-section displays a well-known logarithmic divergence for both maximal and minimal transferred momenta $\sigma \sim \log\left(\frac{q_{max}}{q_{min}}\right)$. The maximal transferred momentum is $q_{max} = 2p$.
The minimal momentum transfer is related to the maximal impact parameter, $\bmax$, to be discussed below.
Using the relation between the scattering angle and the impact parameters in the Schwarzschild metric (for $b \gg r_g = 2 \GN M$)  (see e.g.~\cite{Wald:1984rg})
\begin{equation}
  \label{eq:scatt_angle}
  \sin\frac\theta 2 = \frac{1}{1+\frac{v^2b}{\GN M}} =\frac{1}{1+\frac{2v^2b}{r_g}}
\end{equation}
we can integrate Eq.~\eqref{eq:flip_dif_crosssection} over $\theta$. 
The resulting total helicity-flip cross section for a Dirac fermion is :
\begin{equation}
  \label{eq:final_crosssection}
  \sigma = \frac{(\GN Mm E)^2\pi}{p^4}\log\left(1+\frac{v^2\bmax}{\GN M}\right) 
\end{equation}
In the relativistic limit $E \sim |\vec p| \gg m$, the cross section~\eqref{eq:final_crosssection} behaves as $\left(\frac{m}{E}\right)^2$ as expected. 
In the non-relativistic limit, the cross-section behaves as $v^{-4}$ times the logarithmic term and is independent of mass $m$.The value of the impact parameter $\bmax$ depends on the type of object. We will see below that even using the largest possible $\bmax$ for all objects will not change our conclusion about the survival of the lepton asymmetry. In what follows we will ignore the internal structure of the massive objects, considering the simplest case of scattering on the gravitational center.

\myparagraph{The Majorana case.} 
According to equation (\ref{eq:dirac_final}), the current coupled to the gravitational field $h^{ab}$ is $J_{ab} = {\gamma_a\partial_b}$. The coupling form of the transition matrix $\mathcal{T}$ contains twice more terms in the Majorana case (see e.g. the review \cite{pal_dirac_2011}) : 
\begin{equation}
\label{eq:majorana_coupling_condition2}
	\mathcal{T} \propto h^{ab}\bar{\psi}(p_f)\left[J_{ab} + CJ_{ab}^TC^{-1}\right]\psi(p_i) 
\end{equation}
where $C$ is the charge conjugation operator.
It was shown \cite{Menon:2008wa, Lai:2021tbw} that both terms in parentheses contribute equally to weak-field coupling.
Since the Majorana action is constructed from the real spinors,
$\psi_{\text{M}} = \frac 1{\sqrt 2}(\psi_{\text{D}} + \psi_{\text{D}}^c)$, 
it contains an additional factor of 1/2, which cancels the factor of $2$ in~\eqref{eq:majorana_coupling_condition2}.
Therefore, equation (\ref{eq:final_crosssection}) is also valid for Majorana fermions since both couplings are identical.

\myparagraph{Result.} 
Finally we find the number of helicity-flips, $N_{\flip}$,  that free-streaming neutrinos could have experienced until now:
\begin{multline}
\label{eq:master}
N_{\flip} \equiv \int_0^{z_0}\frac{ dz}{ (1+z)H(z)}\\
\times\int\limits_{M_{\text{min}}}^{M_{\text{max}}}\hskip -1ex dM\frac{dn(z,M)}{dM} v(z) \sigma\bigl(v(z),M\bigr)
\end{multline}
The cross-section $\sigma\bigl(v,M\bigr)$ is given by Eq.~\eqref{eq:final_crosssection};
$v(z) = v_0(1+z)$ is the neutrino velocity at redshift $z$, $v_0$ is the current neutrino velocity, $H(z) = \sqrt{\Omega_\Lambda + \Omega_M(1+z)^3}$ is the Hubble parameter with $\Omega_M = 0.27, \Omega_\Lambda = 0.63$~\cite{Planck:2018vyg}.
The integral over $dz$ is the time that neutrino has traveled in the expanding Universe between initial redshift $z_0$ and today; while the integral over $dM$ computes the scattering rate, accounting for the number density of scattering centers at redshift $0\le z \le z_0$.
The velocity of neutrinos can change while scattering off  the largest objects.
Therefore, to simplify our computations we estimate $N_\flip$ from above by substituting $v(z) \to v_0$ for objects with the mass $M > 10^{14} M_{\odot}$ in the expression for $\sigma\bigl(v,M\bigr)$. 

\begin{figure}[h]
		\includegraphics[width=\linewidth]{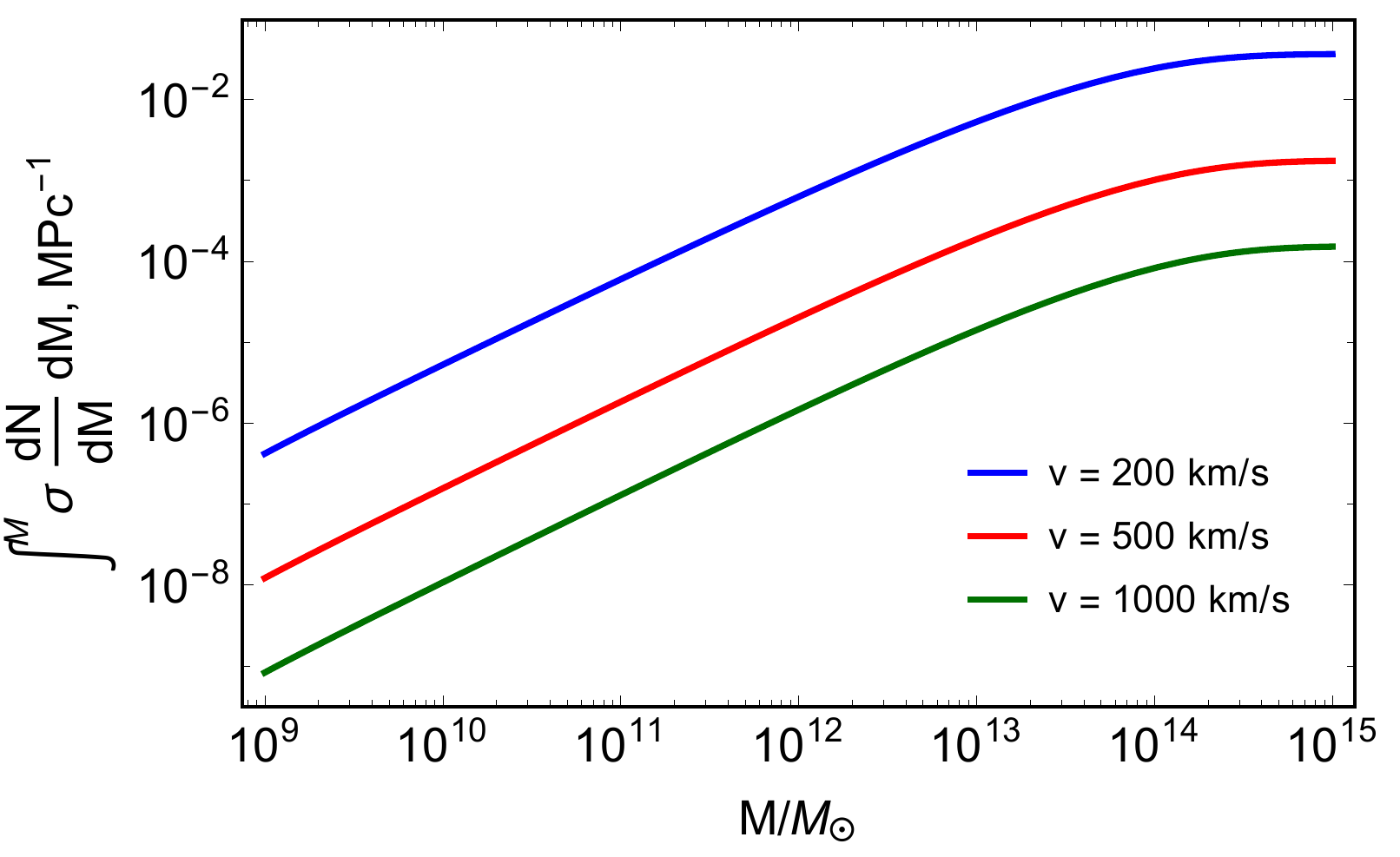}
	\caption{The mass integral of Eq.~\eqref{eq:master} as a function of $M_{\max}$ for various neutrino speeds $v_0$ at redshift $z= 0$. The saturation of the integral appears for masses $M \sim 10^{13}-10^{15} M_{\odot}$ while the contribution from lower masses is negligible.}
	\label{fig:Mdependence_speed}
\end{figure}

To evaluate $\frac{dn(z,M)}{dM}$ we use the Press-Schechter formalism \cite{Press:1974}, see e.g.\ the textbook~\cite{gorbunov_introduction_2011} for necessary details. 
The integral over masses is saturated around $M \sim 10^{14}M_{\odot}$, see Figure~\ref{fig:Mdependence_speed}.
At high masses, $dn/dM$ falls exponentially and the integral converges fast.
Low masses do not contribute significantly due to the $M^2$ dependence of the cross-section~\eqref{eq:final_crosssection}.
This allows us to avoid uncertainties of the Press-Schechter formalism at small masses and therefore we do not revert to more sophisticated methods like e.g.\ \cite{Sheth:1999su}. 
Finally, the integral is dominated by redshifts $z\to 0$ (where the velocities are the smallest and the structures have grown), making the actual choice of $z_0 > 1$ unimportant. 

\begin{figure}[!t]
  \includegraphics[width=\linewidth]{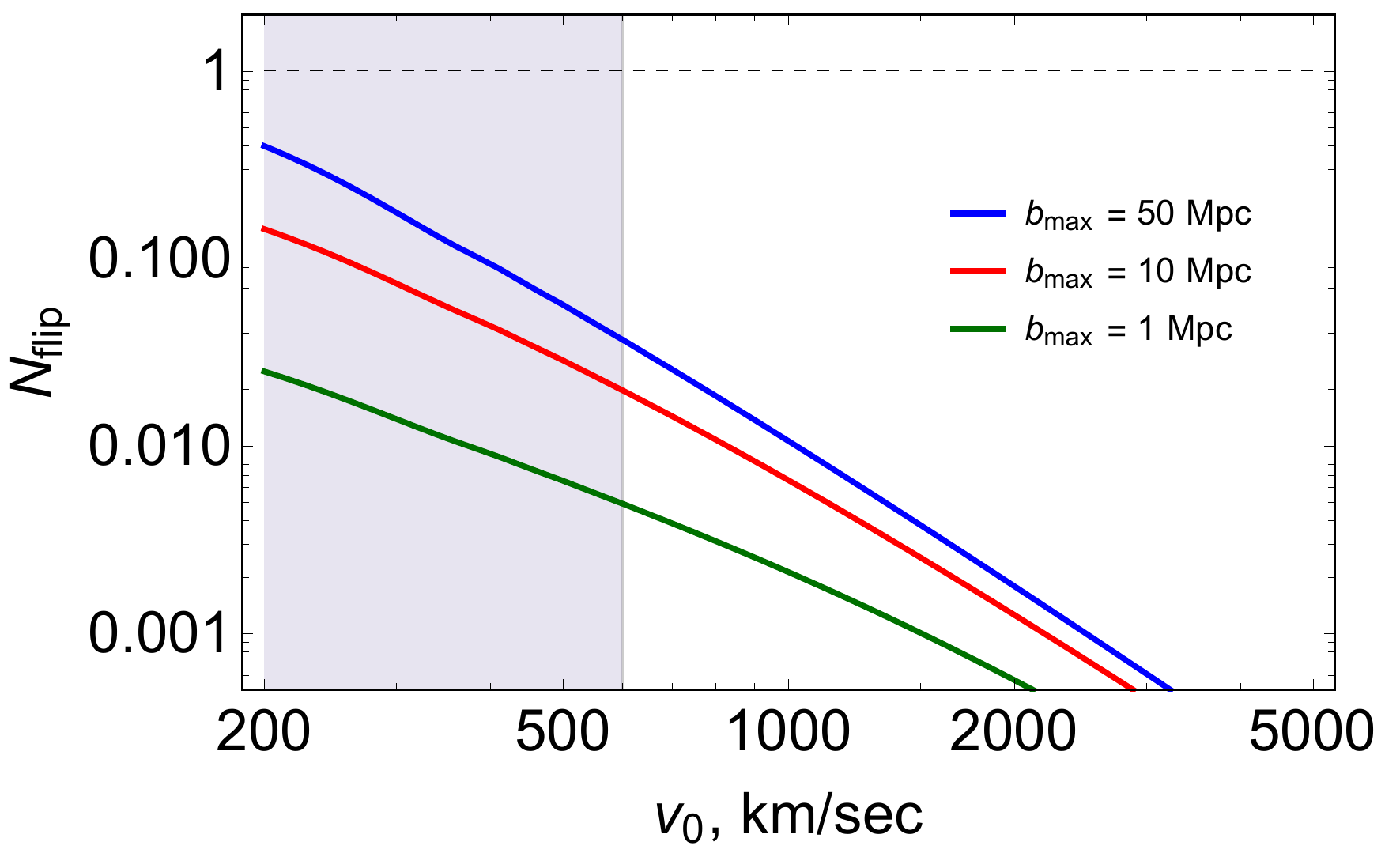}\\
  \includegraphics[width=\linewidth]{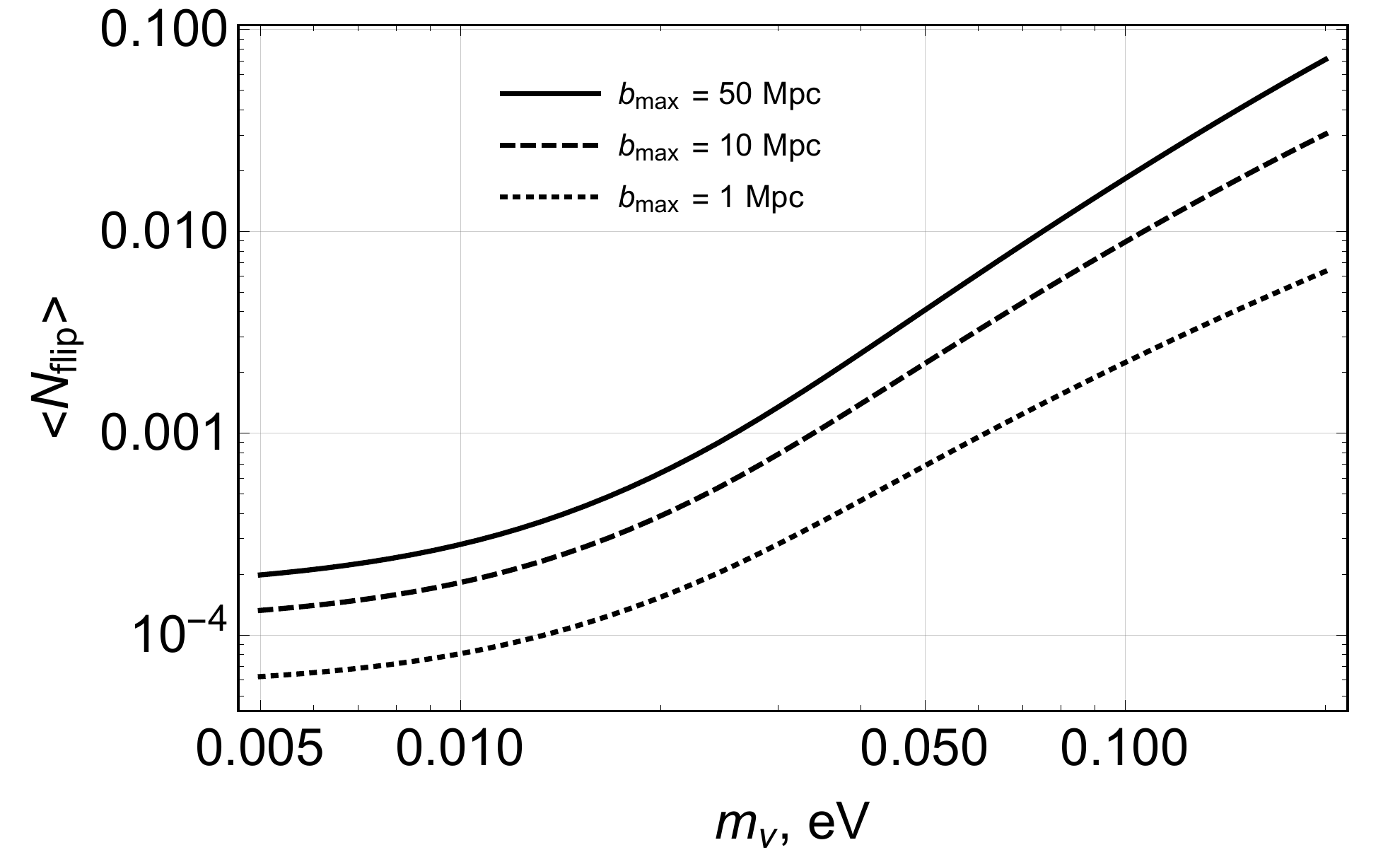}
  \caption{\textbf{Top panel:} The mean number of helicity-flips experienced by free-streaming neutrinos as a function of their velocity \emph{today} (unshaded region).
    The number is significantly smaller than $1$ for all admissible values of $\bmax$.
    The shaded region corresponds to the sub-population of neutrinos, gravitationally bound to the Milky Way and having their helicity erased.
    Note, that the result \emph{does not depend on the mass of neutrinos} as long as they are non-relativistic.\newline \textbf{Bottom panel}: Average $N_\flip$ experienced by the free-streaming neutrinos passing through the Milky Way ($v> v_\esc$) for different values of the impact parameter $b_{\text{bmax}}$. $\langle N_\flip\rangle$ is the quantity in the top panel averaged with the distribution function \protect\eqref{eq:FD}.
    The mass dependence is due to the mass dependence of $f_\nu(v)$. }
	\label{fig:Nflip}
\end{figure}

The final results for $N_\flip$ are presented in Figure~\ref{fig:Nflip} (using $M_{\text{min}} = 10^8M_\odot$, $M_{\text{max}} = \num{2.3e15}M_\odot$ and $z_0 = 5$)\footnote{Larger redshifts can be ignored, since the most massive objects are not formed yet and velocities of neutrinos are too high for effective helicity-flipping scattering.}).
If $N_{\flip} \geq 1$ we consider helicities to be equilibrated, i.e.\ the lepton number erased.
We see that $N_{\flip}$ monotonically decreases with $v_0$, \emph{never reaching} $N_\flip \simeq 1$ for gravitationally unbound neutrinos with $v_0 \ge v_\esc$.
Neutrinos with $v < v_\esc$ are gravitationally bound to the Milky Way and therefore we consider their asymmetry fully erased which changes the total lepton asymmetry by small fraction, as discussed above.
The dependence on $\bmax$ is weak with typical $\bmax$ being about $\sim \CO(\SI{10}{Mpc})$. 
Finally, we stress that $N_\flip$ is independent of the neutrino mass (for non-relativistic neutrinos).
The mass dependence seen in the lower panel of Fig.~\ref{fig:Nflip} is solely due to the bound fraction being dependent on mass (c.f.\ Fig.~\ref{fig:fraction}).

\myparagraph{Conclusion.}
Lepton asymmetry (different numbers of leptons and anti-leptons) can be generated at some early cosmological epochs and be encoded in the cosmic neutrino background (\cnb).
If neutrinos are Majorana particles, this asymmetry is not protected by any conservation law.
Nevertheless \emph{it would remain largely intact today}.
To demonstrate this fact we computed the probability of helicity-flipping gravitational scattering of free-streaming neutrinos and showed that a non-relativistic neutrino would experience $N_\flip \ll 1$ over the lifetime of the Universe.
This conclusion is valid for any neutrino mass as long as neutrinos are non-relativistic today. 
The fraction of free-streaming neutrinos in the Earth's vicinity is estimated to be between $\sim 85\%$ and $\sim 99\%$ for currently admissible values of neutrino masses.
The remaining small fraction of neutrinos are gravitationally bound and their asymmetry is erased. 
If neutrinos are Dirac particles, the total lepton number is, of course, conserved, but is re-distributed between active and sterile sectors.
The above conclusion is then applied to the active sector (left-chiral particles and right-chiral anti-particles).

Our results demonstrate that if the primordial lepton asymmetry had ever been generated, it may in principle be detectable via e.g.\ precise measurements of the neutrino capture rate in Tritium~\cite{PTOLEMY:2018jst} or other elements~\cite{Mikulenko:2021ydo}.
Indeed, the lepton asymmetry changes the neutrino number density and hence the capture rate.
This will of course require percent level precision of measurements (for potential pitfalls see~\cite{Cheipesh:2021fmg,PTOLEMY:2022ldz}).
Additionally, the change of the neutrino capture rate may also be due to the local \cnb overdensity~\cite{deSalas:2019kpa}.
The two scenarios may be distinguished in the case of Dirac neutrinos with negative chemical potential $\mu/T_\nu < 0$.
In this case, the capture rate would also be \emph{lower} than in the standard case -- an effect that cannot be imitated by the overdensity. \footnote{For Majorana neutrinos the rate is proportional to $(\mu/T_\nu)^2$ and always increases, see e.g.\ \cite{Long:2014zva}.}
Confronting such results with the determination of the lepton asymmetry from primordial nucleosynthesis or the cosmic microwave background (see e.g.\ \cite{Escudero:2022okz}) may provide an incredible test of the Big Bang theory.

\begin{acknowledgements}
  \myparagraph{Acknowledgements.}
We thanks M.~Ahlers and M.~Bustamante for a careful reading of the manuscript and useful suggestions.
We also thank A.~Long and M.~Shaposhnikov for comments and suggestions during various stages of this work.
The work received support from the Carlsberg Foundation's grant CF17-0763.  
\end{acknowledgements}

\bibliography{CnuBBIbliography} 
\end{document}